\documentclass[conference]{IEEEtran}
\IEEEoverridecommandlockouts
\usepackage{cite}
\usepackage{bm}
\usepackage{adjustbox}
\usepackage{tabularx,ragged2e,booktabs}
\usepackage{multirow}
\usepackage[cmex10]{amsmath}
\usepackage{makecell}
\usepackage{graphicx}
\usepackage{url}
\usepackage{color}
\usepackage{diagbox} 
\usepackage{amsfonts}
\usepackage{amsmath}
\usepackage{extarrows}
\usepackage{amsfonts}
\usepackage{caption}
\usepackage{algorithm}
\usepackage{algpseudocode}
\usepackage{indentfirst}
\usepackage{caption}
\usepackage{graphics}
\usepackage{booktabs}
\usepackage{epsfig}
\usepackage{soul}
\usepackage{balance} 
\usepackage{caption}
\usepackage{gensymb}
   \def\bI{{\mathbf{I}}}

\def\ba{{\mathbf{a}}}    
\def\bf{{\mathbf{f}}}

\def\bu{{\mathbf{u}}}   \def\bx{{\mathbf{x}}} \def\by{{\mathbf{y}}}
\def\bz{{\mathbf{z}}}

\begin{document}
\title{Decision Triggered Data Transmission and Collection in Industrial Internet of Things }
\author{Jiguang~He$^\dag$,  Long Kong$^\sharp$, Tero~Frondelius$^\ddag$$^\S$, Olli Silv\'en$^\star$, Markku~Juntti$^\dag$\\
        $^\dag$Centre for Wireless Communications, FI-90014, University of Oulu, Finland\\
        $\sharp$Interdisciplinary Centre for Security Reliablility and Trust (SnT), University of Luxembourg, Luxembourg\\
        $^\star$Center for Machine Vision and Signal Analysis (CMVS), FI-90014, University of Oulu, Finland\\
        $^\ddag$R\&D and Engineering, W\"artsil\"a, P.O. Box 244, 65101 Vaasa, Finland\\
        $^\S$University of Oulu, Erkki Koiso-Kanttilan katu 1, 90014 Oulu, Finland\\
 Email: \{jiguang.he, olli.silven, markku.juntti\}@oulu.fi, long.kong@uni.lu, tero.frondelius@wartsila.com\\
        \thanks{This work has been performed in the framework of the IIoT Connectivity for Mechanical Systems (ICONICAL), funded by the Academy of Finland. This work is also partially supported by the Academy of Finland 6Genesis Flagship (grant 318927).}}
 \maketitle
\begin{abstract}
We propose a decision triggered data transmission and collection (DTDTC) protocol for condition monitoring and anomaly detection in the industrial Internet of things (IIoT). In the IIoT, the collection, processing, encoding, and transmission of the sensor readings are usually not for the reconstruction of the original data but for decision making at the fusion center. By moving the decision making process to the local end devices, the amount of data transmission can be significantly reduced, especially when normal signals with positive decisions dominate in the whole life cycle and the fusion center is only interested in collecting the abnormal data. The proposed concept combines compressive sensing, machine learning, data transmission, and joint decision making. The sensor readings are encoded and transmitted to the fusion center only when abnormal signals with negative decisions are detected. All the abnormal signals from the end devices are gathered at the fusion center for a joint decision with feedback messages forwarded to the local actuators. The advantage of such an approach lies in that it can significantly reduce the volume of data to be transmitted through wireless links. Moreover, the introduction of compressive sensing can further reduce the dimension of data tremendously. An exemplary case, i.e., diesel engine condition monitoring, is provided to validate the effectiveness and efficiency of the proposed scheme compared to the conventional ones. 
\end{abstract}


\section{Introduction}
In the foreseeable future, not only all the humans but all the things will be interconnected, entitled as Internet of everything (IoE), with the help of ubiquitous computing and pervasive connectivity. Industrial Internet of things (IIoT), as one key component of IoE, has been bringing great societal impact to the industry~\cite{Wollschlaeger2017}. It is driven by industrial automation and digitalization, and applicable to multifarious application fields, just to name a few, smart manufacturing (to control the manufacturing environment and monitor the production lines), smart city (to use different IoT sensors to collect data and use the insights gained from it to increase operational efficiency), autonomous driving, drone, vessel, and telemedicine~\cite{Sisinni2018,Luo2018}.

In order to support the ever-growing number of connected things, e.g., consumer devices, drones, vehicles, and continuous data transfer, a huge amount of spectrum resources are in high demand. Nevertheless, the problem of spectrum crunch already exists, which is certain to continue as the number of connected devices is expected to grow exponentially. In order to address this critical problem, new methodologies should be considered for the paradigm shift from the conventional human-centric scenarios to future machine-centric ones. In the human-centric applications, the processing and transmission of the data is to reconstruct the original data at the receiver side with a required fidelity. However, for the machine-centric IIoT, the purpose of data collection is usually for extracting the critical features in the data and making real-time decisions~\cite{Xu2014}. Spectrum- and cost-efficient schemes should be adopted to reduce the amount of data to be transmitted while maintaining the accuracy level of the decision making by taking into account the new features of the machine-centric IIoT.

In this paper, we propose a novel concept, which takes advantage of compressive sensing (CS), machine learning (ML), data transmission (DT), and joint decision making \& storage, for efficient data transmission and collection in the IIoT. In the conventional scheme, all the sensor readings are transmitted to the fusion center through wireless links or stored at the end devices. Such a scheme has three drawbacks: 1) The real-time analysis cannot be conducted. 2) It requires a large amount of memory for the end devices. 3) High demand for the channel capacity is required. In order to enable real-time analysis while keeping the memory size reasonably small, our proposed concept can perfectly meet these requirements. On one hand, CS can reduce the dimension of the data to be further processed and transmitted due to the sparsity property of the sensor readings. On the other hand, the module, empowered by the ML techniques, can be leveraged to make local decisions based on the sensed time series. If the anomaly happens less frequently, the normal data is not necessary to be transmitted and stored at the fusion center due to its easy availability. The interesting time series are the abnormal ones, which, on the contrary, should be transmitted, further processed jointly, and stored at the fusion center. We entitle this concept as decision triggered data transmission and collection (DTDTC), especially for IIoT networks. We further provide a case study on condition monitoring and anomaly detection of a big diesel engine in mechanical systems with field-measurement data, where the normal states are supposed to occupy more than $90\%$ of the life cycle of the engine. It is verified that the proposed concept can significantly reduce the amount of data transmission while maintaining the same performance.   
%

\section{Cutting-Edge Techniques}
The IIoT is a cross-disciplinary technology, which includes information and communication technology (ICT), data science, mathematics, and related fields. The promising cutting-edge techniques, which enable the practical implementation of the IIoT, are selected and listed below: 
\begin{itemize}
\item Compressive sensing/sampling for dimension reduction while guaranteeing recovery of the sparse signals with a high probability. 

\item Artificial intelligence (AI), e.g., machine learning (ML), deep learning (DL), and (deep) reinforcement learning (RL), for decision making, prediction, and inference. 

\item Fog computing and edge computing for reducing end-to-end latency.

\item Energy- and cost-efficient communications and networking algorithms for saving power and prolonging network lifespan. 
\end{itemize}

Provided that the sensed signals are smooth or piece-wise smooth\footnote{Most of the cases, signals are smooth or piece-wise smooth in the IIoT, e.g., cylinder pressure discussed in the sequel.}, CS can be leveraged to reduce the dimension of the signals  logarithmically while guaranteeing a satisfactory reconstruction accuracy~\cite{Candes2006}. The low-complexity ML techniques, e.g., one-class support vector machine (SVM)~\cite{Erfani2016} and random decision forests~\cite{Ho1995}, enable decision making at local end devices, mobile edge, fusion center, etc. Edge computing enables computation capability at the edge of any networks, thus, reducing the transmission latency. Energy efficient and near-instant communications and networking schemes, driven by cellular techniques~\cite{Dama2017,Chen2018}, are the state-of-the-art digital transmission techniques for the future IIoT with the guarantee of a wide range of coverage, ultra-high data rate, and ultra-low latency. 

\section{Decision Triggered Data Transmission and Collection}
\begin{figure}[t]
\begin{center}
 \includegraphics[width =7.5cm]{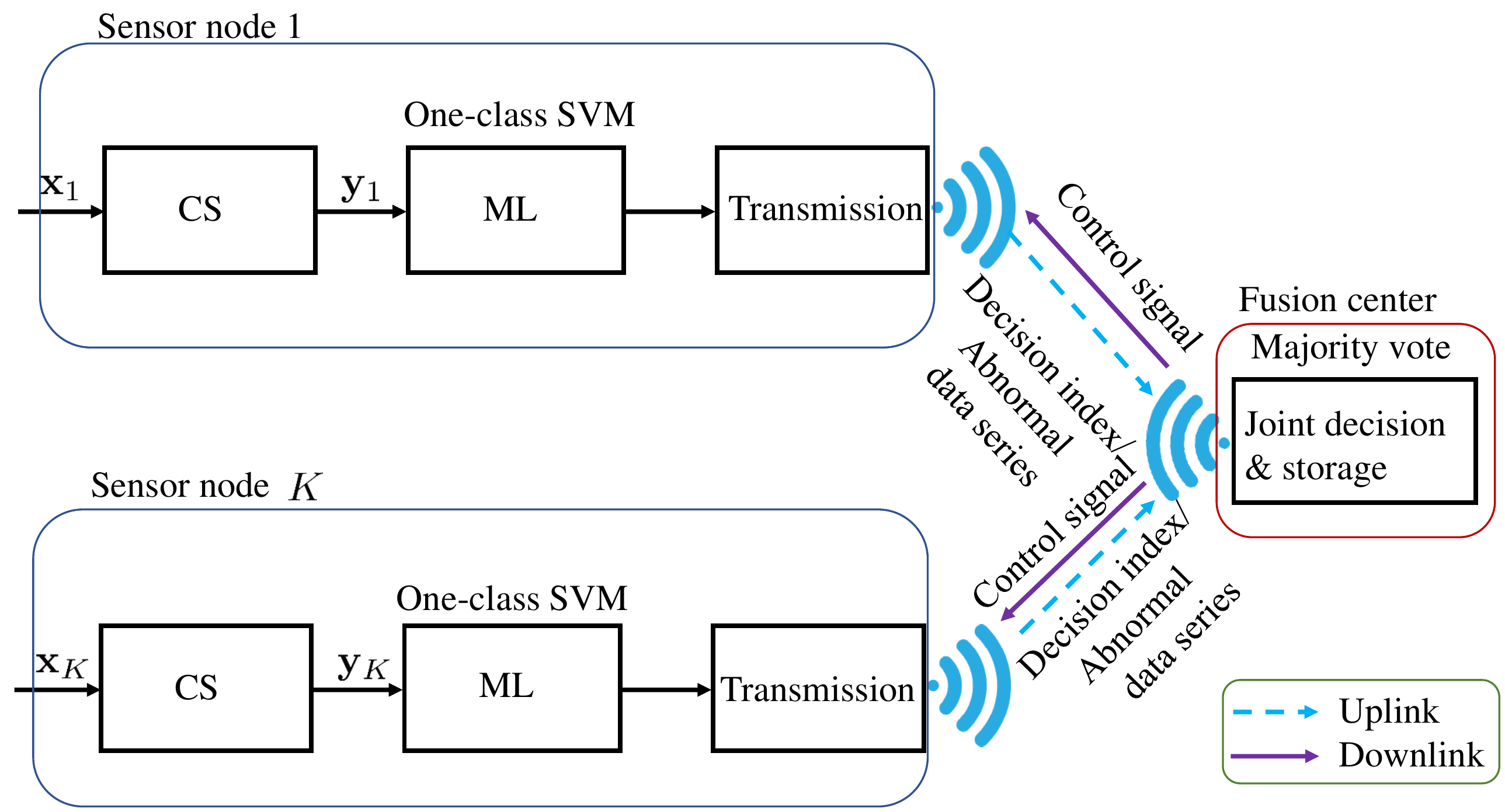}
\end{center}
\caption{Decision triggered data transmission and collection.} \label{DTDTC}
\end{figure}

\textit{Paradigm shift from conventional full data transfer to decision triggered data transmission and collection (i.e., partial data transfer) will play a critical role in the future IIoT networks~\cite{Li2013}.}

In some use cases of the IIoT, most of the sensor readings are not necessary to be transmitted to and stored at the fusion center, e.g., engine state monitoring, due to their easy availability and high-cost transmission. Therefore, we propose the DTDTC concept, shown in Fig.~\ref{DTDTC}, which consists of a CS module, a ML module, and a DT module at each end device (i.e., sensor node), and a joint decision \& storage module at the fusion center, for such use cases.  A binary decision codebook is introduced at the end devices, where ``1'' stands for a normal signal and ``0'' stands for an abnormal signal. Based on the decisions, the end devices determine whether to transmit the decision index or the raw data to the fusion center depending on the decisions. If the decision is positive, i.e., with binary decision ``1'', the end devices notify the index to the fusion center. Otherwise, the abnormal signal is quantized, encoded, and transmitted to the fusion center. After receiving all the signals from the end devices, the fusion center conducts a majority vote for a joint decision. Besides, the fusion center stores the abnormal signals for future usage, e.g., training a more advanced but complicated ML model. We will describe all the modules in details in the sequel. 

\subsection{Compressive Sensing}
Multiple sensor nodes are deployed to sense the same parameter at different locations. Compressive sensing is applied to each original data signals, i.e., $\bx_1,\dots, \bx_K \in \mathbb{R}^{N \times 1}$, where $N$ can be chosen as the number of samples in one period. The source signals are separately sensed through a linear measurement matrix as 
\begin{equation}\label{DCS}
\by_i = \mathbf{\Phi}_i \bx_i + \bz_i, \;\text{for}\; i = 1,\dots, K 
\end{equation}
where each element of $\mathbf{\Phi}_i \in \mathbb{R}^{M_i \times N}$ follows the Gaussian distributions $\mathcal{N}(0, \frac{1}{M_i})$ and $\bz_i\sim\mathcal{N}(\mathbf{0},\sigma_{\bz_i}^2\bI_{M_i})$ is the additive white Gaussian noise (AWGN). We assume that $\mathbf{\Phi}_1, \dots, \mathbf{\Phi}_K$ are fixed and known at the decoder. The source signal $\bx_i = \mathbf{\Psi}\bu_i$ is assumed to have a sparse representation in a transform domain, where $\mathbf{\Psi} \in \mathbb{R}^{N\times N}$ is usually an orthonormal matrix, i.e., $\mathbf{\Psi}^T \mathbf{\Psi} = \bI_N$, with $()^T$ denoting the transpose and $\bI_N$ being the identity matrix with dimension $N\times N$, and $\bu_i$ is the sparse transform coefficient vector. The support of $\bu_i$ is expressed as $\Omega_{\bu_i} = \{j| u_{i,j} \neq 0\}$, and the cardinality of $\Omega_{\bu_i}$ is $C_i$, i.e., $|\Omega_{\bu_i}| = C_i$. Without loss of generality, we set $M_1 = \cdots = M_K = M$. The measurement rate of the CS is defined as $R_\text{M} = \frac{M}{N}$. 
\subsection{One-Class SVM}
In the scenario of condition monitoring, it is easy to get normal sensor readings. In this sense, normal and abnormal training samples are highly unbalanced. The one-class SVM can be used here as a decision making technique. The general idea is to map the input training data to a high dimensional feature space and iteratively find the maximal margin in the hyperplane which best separates the training data from the origin. For the training (using the measurements from~\eqref{DCS}, defined as $\{\by_{\text{T},1},\cdots, \by_{\text{T},m}\}$), the optimization problem is formulated as follows~\cite{Erfani2016}:
\begin{align}
\min_{\ba, R, \boldsymbol{\xi}} R^2 + \frac{1}{m \nu} \sum_{l = 1}^m \xi_l \nonumber\\
\text{s.t.}\; \|\phi(\by_{\text{T},l}) - \ba\|^2 \leq R^2 + \xi_l,\nonumber\\
\forall l = 1,\dots, m,\; \xi_l \geq 0,
\end{align}
where $\ba$ is the center of the hypersphere and its radius is $R$, $m$ is the number of training samples, $\nu$ is the predefined parameter that controls the trade-off between the size of the hypersphere and the fraction of training samples falling outside the hypersphere. Terms $\xi_l, l  = 1, \dots, m$ are the slack variables that allow a portion of the training samples to lie outside the hypersphere. The function $\phi(\cdot)$ is used to map the training samples to a higher dimensional space, e.g., $\phi(\cdot): \mathcal{R}^M \rightarrow \mathcal{R}^Q, M<Q$. For the testing, if the new input $\by$ satisfies $\|\phi(\by) - \ba\|^2 > R^2$, one can claim that an anomaly is detected, and vice versa. 

\subsection{Data Transmission}
In the existing cellular IoT standards, e.g., the extended coverage GSM IoT (EC-GSM-IoT) and narrowband IoT (NB-IoT), a low-complexity channel code (i.e., tail-biting convolutional code (TBCC)) and lower-order modulation are considered~\cite{Liberg2017}. It is reasonable that low-complexity modulation and coding scheme (MCS) are adopted in the IoT end devices due to their limited computation power, memory size, and battery capacity. 

Besides cellular IoT standards, there exist a series of short-range communication standards, such as wireless local area networks (WLANs), for instance, IEEE 802.11~\cite{Tramarin2016}, and wireless personal area networks (WPANs), e.g., IEEE 802.15.1 or IEEE 802.15.4, enabled by bluetooth, Zigbee, etc~\cite{Toscano2012}. The general principle for the MCS is also applied to these standards. In our proposed DTDTC concept, for the index transmission, extremely low-rate short-blocklength channel coding scheme can be considered, while for the abnormal data transmission, we can adopt the scheme: TBCC with lower-order modulation.

\subsection{Joint Decision Making \& Storage}
After receiving all the signals/indices from the end devices, the majority vote is the easiest method for joint decision making at the fusion center based on all the local decisions made at the end devices. The reconstruction of abnormal signals can be done by $\ell_1$-norm based algorithm (e.g., basis pursuit~\cite{Candes2006}), greedy algorithm (e.g., orthogonal matching pursuit (OMP)~\cite{Tropp2007}), and iterative approximate message passing (AMP)~\cite{Donoho2009}. The reconstructed abnormal signals are stored at the fusion center, which will be used for training more advanced machine learning models, e.g., deep neural network (DNN).

\section{Case Study: Engine State Monitoring}
\begin{figure}[t]
\begin{center}
 \includegraphics[width =8.3cm]{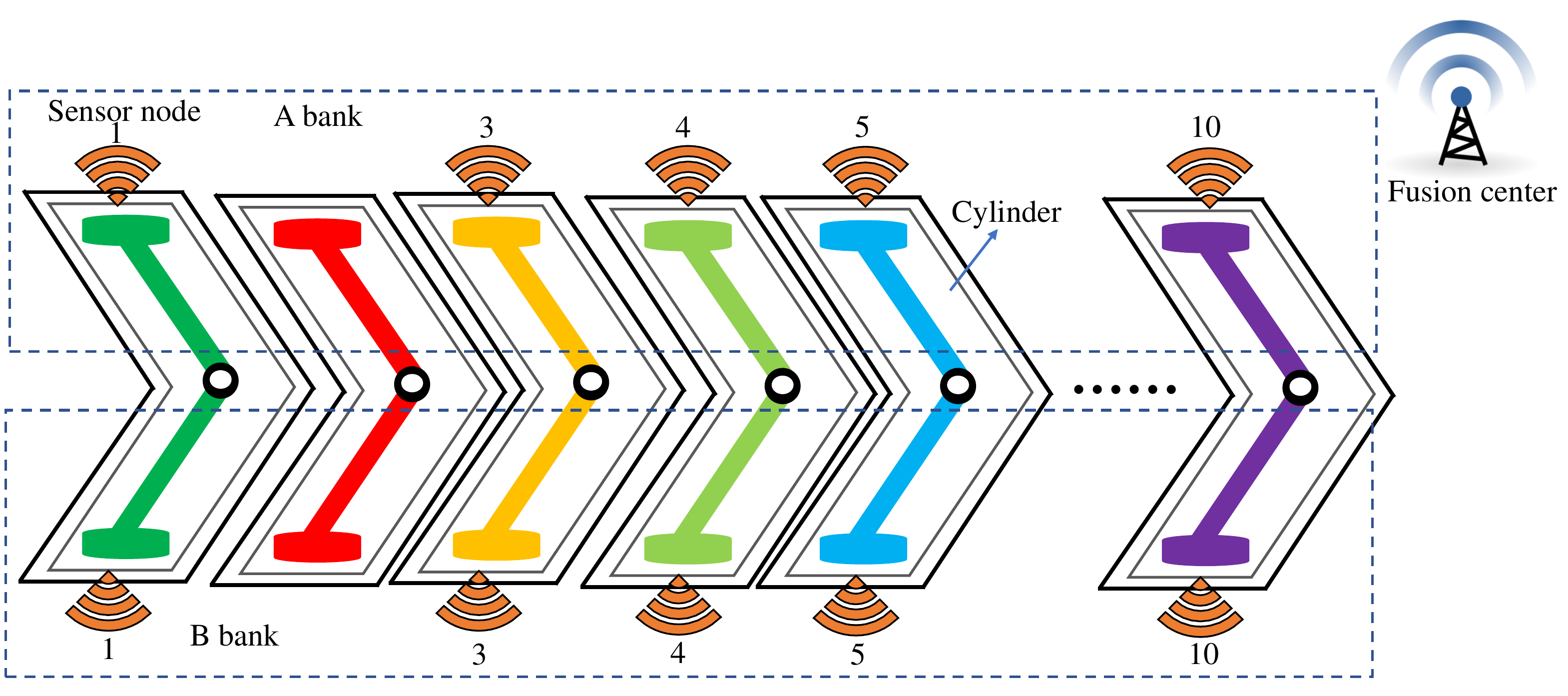}
\end{center}
\caption{A case study on V-configuration diesel engine state monitoring. } \label{Diesel_engine_new}
\end{figure}

\begin{figure}[t]
\begin{center}
 \includegraphics[width =8.3cm]{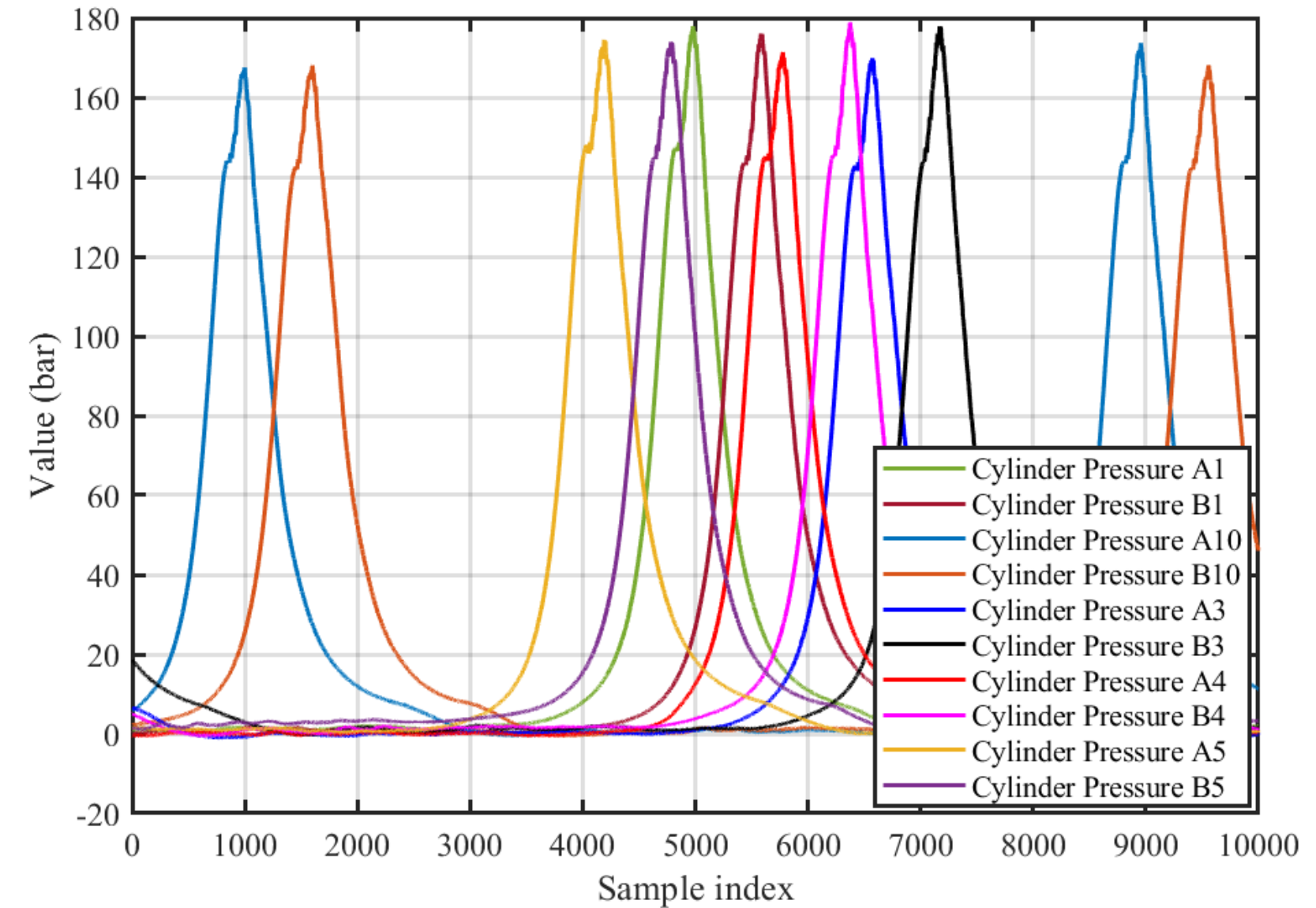}
\end{center}
\caption{A snapshot of cylinder pressure of the diesel engine, provided by W\"artsil\"a.} \label{Cylinder_pressure}
\end{figure}

In this section, we provide a case study on engine state monitoring by following the proposed DTDTC concept. In order to monitor the state of the diesel engine, cylinder pressure is continuously measured by 10 sensor nodes, installed within the V-configuration engine, as shown in Fig.~\ref{Diesel_engine_new}. Each sensor node has the functionalities, depicted in Fig.~\ref{DTDTC}, i.e., CS, ML, and data transmission. In order to provide some intuition about the sensor readings, we provide a snapshot of cylinder pressure from 10 pressure sensors in Fig.~\ref{Cylinder_pressure}.  

Because of the rotation in the mechanical diesel engines, there are good physical intuition that the actual measurements exhibit cyclostationary characteristics~\cite{Antoni2009}. In this sense, for the training and testing of the ML methods, the input of the ML model can be one period of CS measured samples. We follow this principle in the following experiments.

\subsection{Experiments with Field Measurement Signals}
We focus on the field measurement data on cylinder pressure, gathered by $10$ sensors within the engine on 22.2.2013. The sampling rate is $50$ kHz. In order to validate the proposed concept, we run simulations based on half-an-hour samples when the engine is in a normal state. The samples are stored in $30$ H5 files with each file last one minute. That means each file contains $3\times 10^6$ samples. Since the cylinder pressure is cyclostationary, we divide one file into multiple segments with each last one period of time. We use $200$ of them for training one-class SVM, i.e., $m=200$, and the remaining for testing. The performance metrics we consider are the end-to-end accuracy rate (based on joint decision making), transmission efficiency (related to the number of transmitted indices and that of transmitted sequences, including both raw data and compressively sensed data sequences), and reconstruction accuracy for the CS based schemes. For the CS setup, we use random Gaussian measurement matrix with measurement rate set to $0.5$, i.e., $R_\text{M} = 0.5$. In the one-class SVM, we use a linear kernel with $\nu = 0.02$.  We use least absolute shrinkage and selection operator (LASSO)~\cite{bickel2009} for the CS reconstruction at the fusion center with penalizing parameter $\alpha = 0.00001$ and adopt normalized mean square error (NMSE) as the performance metric. The simulation parameters are summarized in Table~\ref{Simulation_parameters}.
\begin{table}
    \centering
    \caption{\textsc{Simulation parameters}}\label{Simulation_parameters}
    \begin{tabular}{|c|c|}
        \hline
        Parameter                 & Value    \\
        \hline
        number of sensors & 10 \\
        sampling rate & 50 kHz \\
        duration      & 30 minutes       \\
        number of H5 files      & 30        \\
        number of training vectors per file        & 200 \\
        number of testing vectors per file      & ~170 \\
        measurement rate   & 0.5      \\
      kernel    & `linear'   \\
      $\nu$    & 0.02   \\
      $\alpha$   & 0.00001      \\
        \hline
    \end{tabular}
\end{table}

In the experiments, it is assumed that the transmission channels are always perfect for the purpose of simplicity. No specific quantization, channel coding, and modulation schemes are taken into consideration. We left these aspects as our future works.
\begin{figure}[t]
\begin{center}
 \includegraphics[width =8.3cm]{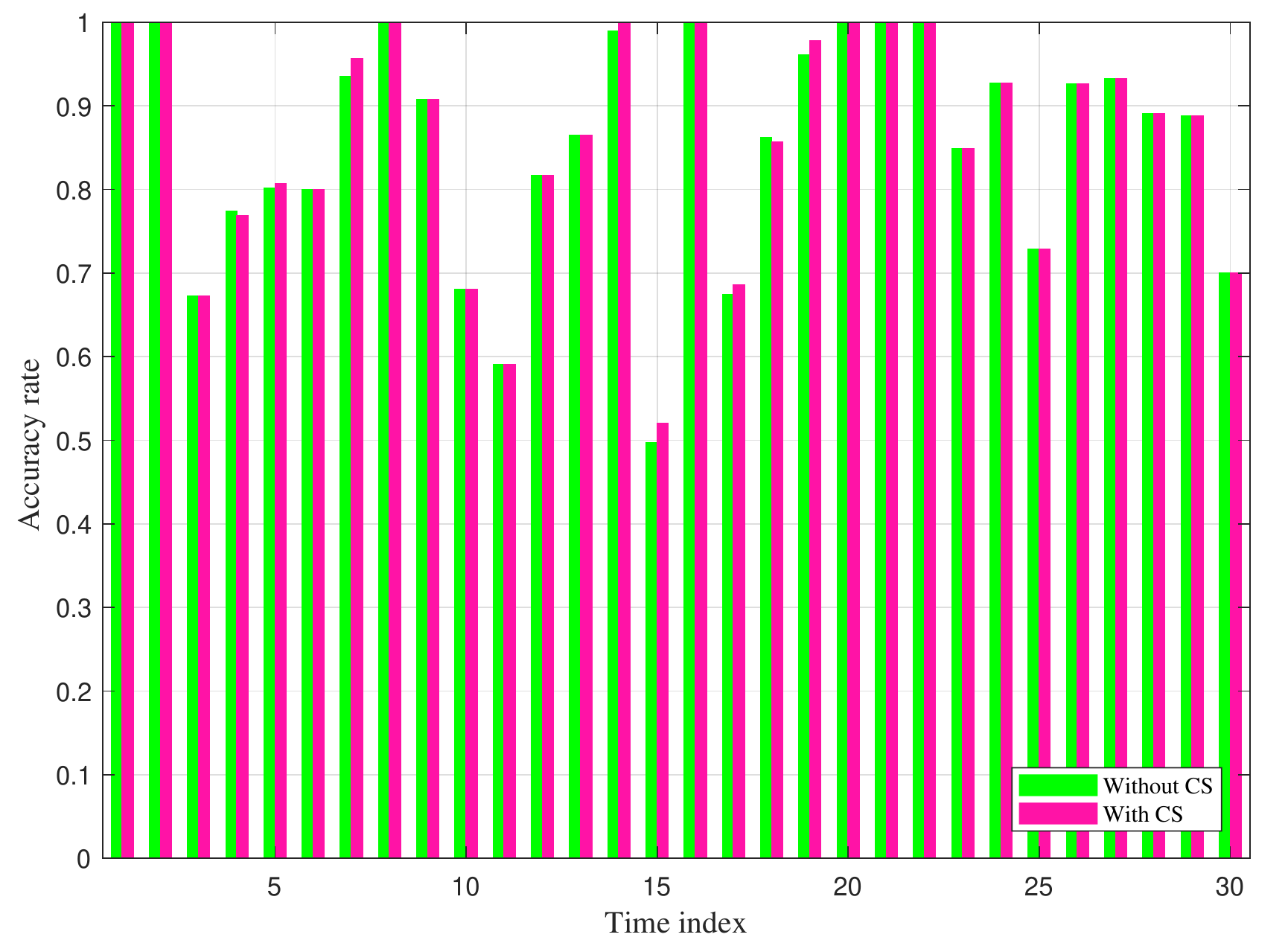}
\end{center}
\caption{End-to-end accuracy rate for the proposed scheme with/without CS.} \label{Accuracy_Rate}
\end{figure}

\begin{table*}[t]
\tiny
\caption{\textsc{Comparison between the proposed DTDTC and the conventional full data transfer scheme.}}\label{Simulation_results}
\scalebox{0.95}{
\begin{adjustbox}{width=1.0\textwidth}
\begin{tabular}{|p{8mm}|p{4mm}|p{9mm}|c|c|c|c|c|c|c|c|c|c|c|c|c|c|c|}
\hline
\multicolumn{3}{|c|}{Time index}&1&2&3&4&5&6&7&8&9&10&11&12&13&14&15\\
\hline
\multirow{ 12}{*}{DTDTC}&\multirow{ 8}{*}{CS}&\makecell{No. of\\ transmitted \\sequences}&23& 5&602&405&459&394&97&0&191&547&687&331&262&84&630\\
\cline{3-18}                                              
&&\makecell{No. of\\ transmitted\\ indices}&1807&1835&1168&1365&1301&1356&1743&1840&1539&1173&1023&1469&1438&1726&1080\\
\cline{3-18}                                                                                                           
&&\makecell{NMSE}&0.0008 &0.0012 &0.0010 &0.0078& 0.0068&0.0120&0.0032 &-&0.0146& 0.0076&0.0021&0.0010&0.0011& 0.0033&0.0048\\ 
\cline{2-18}                                     
&\multirow{ 4}{*}{\makecell{without\\ CS}} &\makecell{No. of\\ transmitted\\ sequences}&24&5&608&404&469&392&115&0&190&548&697&330&262&101&656\\
\cline{3-18}  
&& \makecell{No. of\\ transmitted\\ indices}&1806&1835&1162&1366&1291&1358&1725&1840&1540&1172&1013&1470&1438&1709&1054\\
\hline                                                                                                          
\multirow{ 10}{*}{\makecell{Conventional\\ scheme}}& \multirow{ 4}{*}{ CS} &\makecell{No. of\\ transmitted\\ sequences}&1830&1840&1770&1770&1760&1750&1840&1840&1730&1720&1710&1800&1700&1810&1710\\
\cline{3-18}          
&&\makecell{No. of\\ transmitted\\ indices}&0&0&0&0&0&0&0&0&0&0&0&0&0&0&0\\
\cline{2-18}                                                                                                          
&\multirow{ 4}{*}{\makecell{without\\ CS}} &\makecell{No. of\\ transmitted\\ sequences}&1830&1840&1770&1770&1760&1750&1840&1840&1730&1720&1710&1800&1700&1810&1710\\
\cline{3-18}
&&\makecell{No. of\\ transmitted\\ indices}&0&0&0&0&0&0&0&0&0&0&0&0&0&0&0\\
\hline
\hline

\multicolumn{3}{|c|}{Time index}&16&17&18&19&20&21&22&23&24&25&26&27&28&29&30\\
\hline
\multirow{ 12}{*}{DTDTC}&\multirow{ 8}{*}{CS}&\makecell{No. of\\ transmitted\\ sequences}&0& 443&278&172&1&1&0&280&137&506&145&248&290&243&530\\
\cline{3-18}   
&&\makecell{No. of\\ transmitted\\ indices}&1820&1307&1532&1618&1799&1839&1840&1570&1653&1334&1605&1532&1540&1447&1170\\
\cline{3-18}   
&&\makecell{NMSE}&- &0.0009 &0.0013 &0.0045&0.0019& 0.0019&- &0.0079&0.0029& 0.0026&0.0020&0.0111&0.0099& 0.0068&0.0054\\ 
\cline{2-18}                        
&\multirow{ 4}{*}{\makecell{without\\ CS}} &\makecell{No. of\\ transmitted\\ sequences}&0&460&274&190&2&1&0&281&134&503&146&243&293&250&516\\
\cline{3-18}   
&& \makecell{No. of\\ transmitted\\ indices}&1820&1290&1536&1600&1798&1839&1840&1569&1656&1337&1604&1537&1537&1440&1184\\
\hline               
\multirow{ 10}{*}{\makecell{Conventional\\ scheme}}& \multirow{ 4}{*}{ CS} &\makecell{No. of\\ transmitted\\ sequences}&1820&1750&1810&1790&1800&1840&1840&1850&1790&1840&1750&1780&1830&1690&1700\\
\cline{3-18}          
&&\makecell{No. of\\ transmitted\\ indices}&0&0&0&0&0&0&0&0&0&0&0&0&0&0&0\\
\cline{2-18}                                                                                                          
&\multirow{ 4}{*}{\makecell{without\\ CS}} &\makecell{No. of\\ transmitted\\ sequences}&1820&1750&1810&1790&1800&1840&1840&1850&1790&1840&1750&1780&1830&1690&1700\\
\cline{3-18}                                                                                                    
&&\makecell{No. of\\ transmitted\\ indices}&0&0&0&0&0&0&0&0&0&0&0&0&0&0&0\\
\hline 
\end{tabular}
\end{adjustbox}}
\end{table*}

As shown in Fig.~\ref{Accuracy_Rate}, the simulation results of end-to-end accuracy rate are presented. One can observe that the introduction of CS makes no major difference on the accuracy rate performance compared to the scheme without CS. We also provide the results of the reconstruction accuracy for the scheme with CS\footnote{Note that for some time indices, the values are not provided because the number of transmitted compressively-sensed data sequences is zero.}, the number of transmitted sequences and that of transmitted indices for the schemes with and without CS. Detailed comparisons are made between the proposed DTDTC approach and the conventional transmission scheme (i.e., transmitting all the raw data), shown in Table~\ref{Simulation_results}. The conclusion can be drawn from the results that 1) the introduction of CS does not affect the accuracy of the joint decision making, and 2) our proposed DTDTC can significantly improve the transmission efficiency, since transmitting an index is apparently much more efficient and beneficial for energy saving than transmitting the whole data sequence.

\section{Conclusions and Future Work}

In this paper, we have proposed a novel concept of efficient data transmission and collection (DTDTC) protocol for condition monitoring and anomaly detection. It is feasible and applicable in the IIoT realms by enabling local decision making at the end devices. Our proposed concept has been validated by real-field measurement engine data provided by W\"artsil\"a in terms of reconstruction distortion, accuracy rate of joint decision making, and the number of transmitted indices and sequences. As observed from our numerical discussions, this concept is beneficial for improving energy efficiency, since it can tremendously reduce the amount of transmitted data.

As an initial study, we only use the one-class SVM as the machine learning tool for local decision making. Its simplicity is well tailored for the IoT end devices due to their limited battery and computation capability. There exists a trade-off between the performance and the computational complexity. In the future, more advanced machine learning techniques can be applied when the IoT devices have stronger computation power and are able to harvest energy from their ambient environments.  

Besides, different parameters, including vibration, noise, temperature, stress, etc, should be jointly considered for the condition monitoring and anomaly detection other than only considering one single parameter, e.g., cylinder pressure. 
\balance
\bibliographystyle{IEEEtran}
\bibliography{IEEEabrv,Ref}

\end{document}